\documentclass[preprint,nofootinbib,amsmath,amsfonts,amssymb,onecolumn,superscriptaddress]{revtex4-1}
\pdfoutput=1
\usepackage{graphicx,psfrag,color}% Include figure files
\usepackage{bm}% bold math
\usepackage{mathbbol,verbatim}
\usepackage{slashed}
\usepackage{graphics}
\usepackage{color,ulem}
\allowdisplaybreaks
\usepackage{tikz}
\usetikzlibrary{trees}
\usetikzlibrary{decorations.pathmorphing}
\usetikzlibrary{decorations.markings}
\usetikzlibrary{decorations, decorations.markings, decorations.pathmorphing, arrows, graphs, shapes.geometric, snakes}
\usetikzlibrary{arrows}

\tikzset{
photon/.style={decorate, decoration={snake}, draw=red},
dark/.style={draw=gray, postaction={decorate},
        decoration={markings,mark=at position .55 with {\arrow[draw=gray]{>}}}},
antidark/.style={draw=gray, postaction={decorate},
        decoration={markings,mark=at position .55 with {\arrow[draw=gray]{<}}}},
electron/.style={draw=violet, postaction={decorate},
        decoration={markings,mark=at position .55 with {\arrow[draw=violet]{>}}}},
neutrino/.style={draw,color=violet,thick, postaction={decorate} },
neutrinolight/.style={draw=blue, postaction={decorate} },
quark/.style={draw=blue, postaction={decorate},
        decoration={markings,mark=at position .55 with {\arrow[draw=blue]{>}}}},
antiquark/.style={draw=blue, postaction={decorate},
        decoration={markings,mark=at position .55 with {\arrow[draw=blue]{<}}}},
heavyquark/.style={draw=purple, postaction={decorate},
        decoration={markings,mark=at position .55 with {\arrow[draw=purple]{>}}}},
antiheavyquark/.style={draw=purple, postaction={decorate},
        decoration={markings,mark=at position .55 with {\arrow[draw=purple]{<}}}},
        gluon/.style={decorate, draw=or,
        decoration={coil,amplitude=2pt, segment length=3pt}},
gluon/.style={decorate, draw=or,
        decoration={coil,amplitude=2pt, segment length=3pt}},
ZZ/.style={decorate, decoration={snake,amplitude=1.5pt, segment length=5pt}, draw=greeen},
left,
  }

\definecolor{greeen}{rgb}{0.03,0.84,0.13}
\definecolor{test}{rgb}{0.03,0.74,0.33}
\definecolor{viol}{rgb}{0.44,0,0.94}
\definecolor{or}{rgb}{0.95,0.65,0}

\begin{document}

\preprint{UMD-PP-017-30}

\title{ Same Sign versus Opposite Sign Dileptons as a Probe of Low Scale Seesaw Mechanisms}

\author{Arindam Das}
\affiliation{School of Physics, KIAS, Seoul 130-722, Korea}
\author{P. S. Bhupal Dev}
\affiliation{Department of Physics and McDonnell Center for the Space Sciences,  Washington University, St. Louis, MO 63130, USA}
\author{Rabindra N. Mohapatra}
\affiliation{Maryland Center for Fundamental Physics, Department of Physics, University of Maryland, College Park, MD 20742, USA}

%\emailAdd{arindam@kias.re.kr}
%\emailAdd{bdev@wustl.edu}
%\emailAdd{rmohapat@umd.edu}

\begin{abstract}
We calculate the ratio $R_{\ell\ell}$ of same sign (SS) to opposite sign (OS) dileptons in type I and generalized inverse seesaw models and show that it can be anywhere between 0 and 1 depending on the detailed texture of the right-handed neutrino mass matrix. Measurement of  $R_{\ell\ell}$ in hadron colliders can therefore provide a way to probe the nature of seesaw mechanism and also to distinguish between the two types of seesaw mechanisms.  We work within the framework of left-right symmetric model as an example. We emphasize that coherence of the final states in the $W_R$ decay is crucial for this discussion and it requires the right-handed neutrinos to be  highly degenerate. We isolate the range of parameters in the model where this effect is observable at the LHC and future colliders.
\end{abstract}

%\keywords{Neutrino Mass, Seesaw mechanism, Large Hadron Collider}
 
\maketitle

\section{Introduction}  Different kinds of seesaw mechanism have been proposed as ultraviolet (UV)-complete theories that lead to the dimension-5 Weinberg operator~\cite{weinberg} for understanding small neutrino masses. Two of them are the so called type-I~\cite{seesaw1, seesaw2, seesaw3, seesaw4, seesaw5} and inverse seesaw~\cite{ISS1, ISS2}, which have been widely discussed in the literature. The type-I seesaw involves adding SM-singlet heavy fermions $N$ with Majorana masses that violate lepton number maximally, whereas in the inverse seesaw,  one adds two SM-singlet heavy neutrinos $N$ and $S$ and a small $L$-violating mass for one set of the new singlet fermions.  A simple UV-complete extension of the Standard Model (SM) that incorporates all the key ingredients of both type I seesaw and inverse seesaw and leads to them naturally is the left-right symmetric model~\cite{LR1, LR2, LR3}.  No extra symmetries need to be added to generate the right texture for getting tiny neutrino masses. The right-handed neutrino (RHN), predicted by anomaly considerations in this theory, couples to the right-handed (RH) gauge boson $W_R$ and is the source of the lepton number violating (LNV) signal~\cite{KS} we will discuss. In this paper we will work within the framework of the minimal left-right model and assume that $W_R$ is kinematically accessible to the colliders. In other words, for $\sqrt s=14$ TeV LHC, we assume the mass of the $W_R$ boson to be less than 5 TeV or so~\cite{Ferrari:2000sp, Deppisch:2015qwa}. 

A key predictions for the TeV-scale left-right type-I seesaw model is that it leads to a spectacular LNV signal in hadron colliders in the form of two same-sign leptons and two jets with no missing energy~\cite{KS}.  This arises from the production and decay of heavy RHNs, both mediated by the $W_R$ gauge boson in the $s$-channel. The Majorana nature of the RHN dictates that the final states with same-sign (SS) dileptons ($\ell^\pm\ell^\pm$) appear in equal number with opposite-sign (OS) dilepton states ($\ell^\pm\ell^\mp$). In other words, the minimal left-right type-I seesaw prediction  is that the ratio of the number of events in the two final states, 
$R_{\ell\ell}\equiv N_{\rm SS}/N_{\rm OS}=1$. 
This in fact is considered a `smoking gun' signal for TeV-scale type-I seesaw in general\footnote{The minimal TeV-scale type-I seesaw (without any additional gauge or Higgs interactions) requires large light-heavy neutrino mixing in order to have an observable signal at colliders~\cite{Deppisch:2015qwa, Atre:2009rg, Antusch:2016ejd}.} and, more specifically, for the left-right seesaw model and has been extensively studied in the literature, both for the LHC~\cite{KS, Ferrari:2000sp, Deppisch:2015qwa, Nemevsek:2011hz, Das:2012ii, AguilarSaavedra:2012gf, Han:2012vk, Chen:2013fna, Khachatryan:2014dka, Ng:2015hba, Dev:2015kca, Gluza:2016qqv,  Mitra:2016kov, Ruiz:2017nip,  Aad:2015xaa, Dev:2016dja, Roitgrund:2017byx}, as well as other future colliders~\cite{Lindner:2016lxq, Mondal:2015zba, Biswal:2017nfl, Golling:2016gvc}.  %Coloma:2015una,Dobrescu:2015qna,Dev:2016dja,Dobrescu:2015yba,Dobrescu:2015jvn,Deppisch:2015cua}

On the other hand, in the inverse seesaw mechanism, lepton number breaking is very small, because the heavy singlet neutrino ($N$) is paired with another singlet fermion ($S$) to form a pseudo-Dirac pair and the Majorana nature of the neutrino emerges from a keV-scale Majorana mass $\mu_S$ of $S$ fermion (for TeV-scale seesaw). This model when
 embedded into the TeV-scale left-right framework exhibits some interesting features. There are two versions of this model: the minimal version where there is no majorana mass for the $N$~\cite{Dev:2009aw, An:2011uq, Chen:2011hc} and a second more general one where there is a Majorana mass $\mu_R$ for $N$~\cite{Dev:2012sg, Awasthi:2013ff, Dev:2015pga}. In the minimal version, the leading order prediction for collider signal is that final states will approximately conserve lepton number, implying that $R_{\ell\ell}\simeq 0$~\cite{Chen:2011hc}.  In the more general inverse seesaw, which can also arise from left-right seesaw models~\cite{Dev:2015pga}, the neutrino mass formula remains unaffected at the tree-level, although there is an unavoidable one-loop contribution from electroweak radiative corrections~\cite{Dev:2012sg}; however the $N$ fermion has a potentially large Majorana mass that breaks lepton number by two units. The question remains as to how do the dilepton final states look like in this general case i.e. is $R_{\ell\ell}=1$ or different?  This question has been recently studied in some special cases~\cite{Dev:2015pga, Anamiati:2016uxp, Antusch:2017ebe} and was shown that due to interference between two heavy Majorana neutrino mass eigenstates, one could in principle realize a scenario with $R_{\ell\ell}$ anywhere between 0 and 1. The goal of this study is to do a more general analysis and discuss whether analyzing dilepton states in a hadron collider via production of a $W_R$ boson, one can probe the details of the RHN mass matrix and distinguish between the type-I and general inverse seesaw mechanisms. 
 
 The rest of the paper is organized as follows. In Section~\ref{sec:coh} we discuss the coherence condition for interference between two heavy Majorana neutrino mass eigenstates, which plays a crucial role in our discussion.  In Section~\ref{sec:typeI}, we apply the coherence conditions to discuss the nature of dilepton final states in type-I seesaw. In Section~\ref{sec:inv} we explain the general inverse seesaw model. In Section~\ref{sec:inv2} we apply the coherence conditions for the inverse seesaw case to get the $R_{\ell\ell}$ as a function of parameters of inverse seesaw model. We give our conclusions in Section~\ref{sec:con}. Some useful three-body decay widths for the RHN are listed in Appendix~\ref{sec:app}. 
 
 \section{Coherence Conditions for Interference} \label{sec:coh}
When  a $W_R$ gauge boson is produced in proton-proton collisions, it decays into flavor eigenstates of the RHNs $N_{\ell}$ along with the corresponding charged lepton $\ell_R$ (where $\ell=e,\mu,\tau$). For simplicity, let us consider two RHNs, say $N_e$ and $N_\mu$. When these flavor eigenstates evolve, they do so as linear combination of mass eigenstates $N_{1,2}$. The $N_{1,2}$ are linear combinations of $N_e$ and $N_\mu$ in the type-I seesaw case and of $N$ and $S$ in the inverse seesaw case. The $N_{1,2}$ are Majorana fermions and they will evolve and interfere as they produce the charged leptons (along with two jets) in their final state. Only if the coherence condition (discussed below) is satisfied, they will interfere; otherwise they will simply give equal number of SS and OS dilepton final states. 
 
 The coherence conditions for light neutrinos have been discussed inRefs.~\cite{Kayser:1981ye, Akhmedov:2007fk}. There are two conditions that must be satisfied for interference between the two states to take place: (i) coherence in emission and (ii) the coherence must be maintained till the RHNs decay i.e. for their full decay length. The results imply that the first condition is satisfied when the uncertainty in their mass square exceeds their actual mass difference. We now transplant their argument to the case of two RHNs at hand. Denoting by $\sigma_{m^2}$ the mass uncertainty, we get for the coherence condition $\sigma_{m^2}$ $\geq \Delta M^2\equiv |M^2_1-M^2_2|$. The $\sigma_{m^2}$ in this case is estimated to be $2\sqrt{2} E\Gamma_{W_R}$ where $E$ is the average energy of the RHN eigenstates and $\Gamma_{W_R}$ is the width of the $W_R$ which causes the uncertainly in the  energy of the produced heavy neutrino state. Thus, in our case, coherence in emission occurs when 
\begin{align}
\Delta M^2 \ \leq \ 2\sqrt{2}E\Gamma_{W_R} \, .
\label{eq:coh1}
\end{align}
 For TeV-scale $W_R$ and RHNs, this is satisfied  when the mass difference between the states is less than few hundred GeV, where we have estimated $\Gamma_{W_R}\simeq (g^2/12\pi) M_{W_R}$, setting the $SU(2)_L$ and $SU(2)_R$ couplings to be equal, i.e. $g_L=g_R\equiv g$. 

Turning to the second condition, we take the decay length $L$ as $L=1/\Gamma_N$ and using the results of Ref.~\cite{Akhmedov:2007fk}, require that $L\leq \sigma_x/\delta v_g$, 
where $\sigma_x$ is the size of the RHN wave packets and $\delta v_g$ is the difference between the group velocities of the individual RHNs. We have 
$\sigma_x \sim ( \sigma_E )^{-1}\sim (\Delta M^2/2\sqrt{2}E)^{-1}$ and $\delta v_g\sim (\Delta M^2/2E^2)$. Putting theses together, we get
\begin{eqnarray}
L \ \equiv \ \frac{1}{\Gamma_N}  \ < \ \frac{4\sqrt{2}E^3}{(\Delta M^2)^2} \, .
\label{eq:coh2}
\end{eqnarray}
This implies a stringent condition on the mass difference between the two interfering RHNs. For instance, for $M_{W_R}=5$ TeV, $M_{N}\simeq 1$ TeV,  and $E\sim 2$ TeV, we get the coherence condition 
$\Delta M\equiv |M_1-M_2|\leq $ GeV. Note that this condition is more stringent than what condition (i) alone would have implied [cf. Eq.~\eqref{eq:coh1}] and requires a degeneracy of one part in $10^3$ between the two RHN masses for interference to take place. In deriving this, we have used the decay width formula for the RHNs given in Appendix~\ref{sec:app}. 
 
 From this discussion, we conclude that if interference effect is observed, it will imply constraints on the mass matrix of both the type I and inverse seesaw, helping to further elucidate the nature of the seesaw. It will for example imply that there are at least two nearly degenerate RHN states, consistent with the general expectation from many TeV-scale seesaw models~\cite{Pilaftsis:1991ug, Gluza:2002vs, Kersten:2007vk, Xing:2009in, He:2009ua, Adhikari:2010yt, Ibarra:2010xw, Mitra:2011qr, Dev:2013oxa, Chattopadhyay:2017zvs}, which require the quasi-degeneracy to satisfy the neutrino oscillation data. This is also the requirement for successful resonant leptogenesis via the out-of-equilibrium decay of TeV scale RHNs~\cite{Pilaftsis:2003gt, Dev:2017wwc}.

 \section{Same Sign vs Opposite Sign Dilepton Events in Type-I seesaw}\label{sec:typeI}
 Let us first briefly recapitulate the well known field theoretic argument of why for Majorana RHNs the final states in its decay have equal number of both sign leptons. For concreteness, we illustrate this in the context of left-right model but the argument is general. In the left-right model, the decay of $N$ can be assumed to occur via the emission of a virtual $W_R$ boson and it comes from the RH gauge interaction
 \begin{eqnarray}
 {\cal L}_I \ = \ \frac{g}{\sqrt{2}}\bar{\ell}_R\gamma_\mu N W^{-,\mu}_R+\frac{g}{\sqrt{2}}  N^TC^{-1}\gamma_\mu\ell_RW^{+,\mu}_R
\label{eq:1}
\end{eqnarray}
The second term in the above equation is nothing but the hermitian conjugate of the first one after we use the Majorana condition for $N$ i.e. $N=C\bar{N}^T$ (where $C$ is the charge conjugation operator). Now note that in both terms the $N$ field is annihilated but the final state from the first  term is an $\ell^-$ whereas that from the second term is an $\ell^+$, while both the amplitudes are the same i.e. $g/\sqrt{2}$. This is the basic reason for equal number of SS and OS dileptons in the final states which for a $pp$ collision leads to their ratio $R_{\ell\ell}=1$.

To see how interference between two RHN states affects the ratio $R_{\ell\ell}$, let us consider the simple case of type-I seesaw with only two heavy neutrino flavors $(N_e,N_\mu)$. This case has been discussed in some details in Refs.~\cite{Bray:2007ru, Gluza:2015goa,  Carmona:2016oxx, Gluza:2016qqv}. Here we emphasize the importance of the coherence condition and present new analytic results on the effect on different flavor combinations of the final states. One can easily generalize this to more flavors, but the main conclusion of this section remains unchanged.  

Including the effect of $CP$ violation, we can write the flavor eigenstates as the following combinations of the mass eigenstates: 
\begin{eqnarray}
N_e \ & = & \ c_\theta N_1+s_\theta e^{i\delta}N_2\, , \nonumber \\
N_\mu\ & = & \ -s_\theta N_1+c_\theta e^{i \delta}N_2 \, ,
\label{eq:3.2}
\end{eqnarray}
where $\delta$ is the $CP$ phase in the RHN mixing, $\theta$ is the mixing angle in this sector, and $c_\theta\equiv \cos\theta,\, s_\theta \equiv \sin\theta$. For the general $2\times 2$ RHN mass matrix 
\begin{align}
{\cal M}_N \ = \ \begin{pmatrix} M_1 & Me^{i\phi} \\ Me^{i\phi} & M_2 \end{pmatrix}\, ,
\label{mass}
\end{align}
the mixing angle is given by 
\begin{align}
\theta \ = \ \frac{1}{2}\tan^{-1}\left|\frac{2M}{M_1-M_2}\right| \, .
\label{eq:theta}
\end{align}
Substituting Eqs.~\eqref{eq:3.2} in the interaction Lagrangian for RHNs in Eq.~\eqref{eq:1}, we get  
 \begin{eqnarray}
 \mathcal{L}_I &\ = \ &\frac{g}{\sqrt{2}}\Big[\bar{e}_R\gamma_\mu (c_\theta N_1+s_\theta e^{i\delta}N_2)W^{-,\mu}_R+(c_\theta N_1+s_\theta e^{-i\delta}N_2)^TC^{-1}\gamma_\mu e_RW^{+,\mu}_R \nonumber \\
 &+&\bar{\mu}_R\gamma_\mu (-s_\theta N_1+c_\theta e^{i\delta}N_2)W^{-,\mu}_R+ (-s_\theta N_1+c_\theta e^{-i\delta}N_2)^TC^{-1}\gamma_\mu\mu_R W^{+,\mu}_R \Big]
 \end{eqnarray}
where we have assumed that RH charged leptons are the mass eigenstate.

Using the coherence conditions, we can write the time evolution of the amplitudes for SS and OS final states as follows:
\begin{eqnarray}
A_{{\rm OS},ee}(t)&\ = \ &c^2_\theta e^{-iE_1 t -\frac{\Gamma_1t}{2}}+s^2_\theta e^{-iE_2 t -\frac{\Gamma_2t}{2}} \, , \label{eq:OSee} \\
A_{{\rm SS},ee}(t)&\ = \ &c^2_\theta e^{-iE_1 t -\frac{\Gamma_1t}{2}}+s^2_\theta e^{-2i\delta} e^{-iE_2 t -\frac{\Gamma_2t}{2}} \, ,\label{eq:SSee} \\
A_{{\rm OS},\mu\mu}(t)&\ = \ &s^2_\theta e^{-iE_1 t -\frac{\Gamma_1t}{2}}+c^2_\theta e^{-iE_2 t -\frac{\Gamma_2t}{2}} \, ,\label{eq:OSmumu} \\
A_{{\rm SS},\mu\mu}(t)&\ = \ &s^2_\theta e^{-iE_1 t -\frac{\Gamma_1t}{2}}+c^2_\theta e^{-2i\delta}e^{-iE_2 t -\frac{\Gamma_2t}{2}} \, , \label{eq:SSmumu}\\
A_{{\rm OS},e\mu}(t)&=&-c_\theta s_\theta \Big[ e^{-iE_1 t -\frac{\Gamma_1t}{2}}- e^{-iE_2 t -\frac{\Gamma_2t}{2}}\Big] \ =\ A_{{\rm OS},\mu e}(t)\, , \label{eq:OSemu} \\
A_{{\rm SS},e\mu}(t)&\ = \ &-c_\theta s_\theta \Big[ e^{-iE_1 t -\frac{\Gamma_1t}{2}}- e^{-2i\delta}e^{-iE_2 t -\frac{\Gamma_2t}{2}}\Big]\ = \ A_{{\rm SS},\mu e} (t) \, ,\label{eq:SSemu}
\end{eqnarray}
where $\Gamma_{1,2}$ are the total decay widths of the two mass eigenstates $N_{1,2}$. 

We adopt the following procedure to get the ratio of SS and OS final states~\cite{Anamiati:2016uxp}:
\begin{eqnarray} 
R_{\ell\ell} \ = \ \frac{\int^\infty_0 dt\left |A_{{\rm SS},\ell\ell}(t)\right|^2}{\int^\infty_0 dt\left |A_{{\rm OS},\ell\ell}(t)\right|^2}  \ \equiv \ \frac{N_{{\rm SS},\ell\ell}}{N_{{\rm OS},\ell\ell}} \, .
\label{Rll}
\end{eqnarray}
In order to illustrate the effect of the interference between the two states, we make the simplifying assumption that the two RHNs are non-relativistic (which is a good approximation when the $W_R$ mass is slightly larger than two times the RHN mass) and approximate $E_{1,2}\simeq M_{N_{1,2}}\simeq M_N\pm \Delta M/2$, where $\Delta M\equiv M_{N_1}-M_{N_2}$ is the mass splitting between the two RHN mass eigenstates. The eigenvalues $M_{N_{1,2}}$ can be obtained by calculating the eigenvalues of Eq.~\eqref{mass}. Then from Eqs.~\eqref{eq:OSee} and \eqref{eq:SSee}, the number of SS and OS dielectron events are respectively  given by 
\begin{eqnarray}
N_{{\rm OS}, ee} & \ = \ & \Gamma_{\rm avg}\left[\frac{c^4_\theta}{\Gamma_1}+\frac{s^4_\theta}{\Gamma_2}+c^2_\theta s^2_\theta\frac{\Gamma_1+\Gamma_2}{\left(\frac{\Gamma_1+\Gamma_2}{2}\right)^2+(\Delta M)^2}\right] \, ,\\ 
N_{{\rm SS},ee} & \ = \ & \Gamma_{\rm avg}\left[\frac{c^4_\theta}{\Gamma_1}+\frac{s^4_\theta}{\Gamma_2}+c^2_\theta s^2_\theta\left\{\frac{(\Gamma_1+\Gamma_2)\cos{2\delta}}{\left(\frac{\Gamma_1+\Gamma_2}{2}\right)^2+(\Delta M)^2}-\frac{2\Delta M \sin{2\delta} }{\left(\frac{\Gamma_1+\Gamma_2}{2}\right)^2+(\Delta M)^2}\right\}\right] \, ,
\end{eqnarray}
where $\Gamma_{\rm avg}\equiv (\Gamma_1+\Gamma_2)/2$. 
Similarly, for dimuon events, we have from Eqs.~\eqref{eq:OSmumu} and \eqref{eq:SSmumu} respectively, 
\begin{eqnarray}
N_{{\rm OS},\mu\mu} & \ = \ & \Gamma_{\rm avg}\left[\frac{s^4_\theta}{\Gamma_1}+\frac{c^4_\theta}{\Gamma_2}+c^2_\theta s^2_\theta\frac{\Gamma_1+\Gamma_2}{\left(\frac{\Gamma_1+\Gamma_2}{2}\right)^2+(\Delta M)^2}\right] \, , \\ 
N_{{\rm SS},\mu\mu} & \ = \ & \Gamma_{\rm avg}\left[\frac{s^4_\theta}{\Gamma_1}+\frac{c^4_\theta}{\Gamma_2}+c^2_\theta s^2_\theta\left\{\frac{(\Gamma_1+\Gamma_2)\cos{2\delta}}{\left(\frac{\Gamma_1+\Gamma_2}{2}\right)^2+(\Delta M)^2}-\frac{2\Delta M\sin{2\delta} }{\left(\frac{\Gamma_1+\Gamma_2}{2}\right)^2+(\Delta M)^2}\right\}\right] \, .
\end{eqnarray}
Finally, for the $e\mu$ events, we have from Eqs.~\eqref{eq:OSemu} and \eqref{eq:SSemu} respectively
\begin{eqnarray}
N_{{\rm OS},e\mu} & \ = \ & N_{{\rm OS}, \mu e} \ = \ \Gamma_{\rm avg} \: c^2_\theta s^2_{\theta}\left[\frac{1}{\Gamma_1}+\frac{1}{\Gamma_2}-\frac{\Gamma_1+\Gamma_2}{\left(\frac{\Gamma_1+\Gamma_2}{2}\right)^2+(\Delta M)^2}\right] \, ,\\ \
N_{{\rm SS},e\mu}& \ = \ & N_{{\rm SS}, \mu e} \ = \ \Gamma_{\rm avg} \: c^2_\theta s^2_{\theta}\left[\frac{1}{\Gamma_1}+\frac{1}{\Gamma_2}-\left\{\frac{(\Gamma_1+\Gamma_2)\cos{2\delta}}{\left(\frac{\Gamma_1+\Gamma_2}{2}\right)^2+(\Delta M)^2}-\frac{2\Delta M\sin{2\delta} }{\left(\frac{\Gamma_1+\Gamma_2}{2}\right)^2+(\Delta M)^2}\right\}\right] \, .\nonumber\\
\end{eqnarray}
%%%%%%%%%%%%%%%%%%%%% FIGURE %%%%%%%%%%%%%%%%%%%%%%%%%%%%%%%%%

%\begin{figure}[t]
%\centering
%\includegraphics[width = 0.75\textwidth]{type1mm.pdf}
%\caption{The variation of $R_{\mu\mu}$ as a function of the $CP$ phase $\delta$ for different values of $\Delta M/\Gamma_{\rm avg}$ in the TeV-scale left-right type-I seesaw model. Here we have chosen a fixed value of $M_{W_R}=5$ TeV and $M_N=500$ GeV.} 
%\label{fig:Rmumu}
%\end{figure}
%%%%%%%%%%%%%%%%%%%%%%%%%%%%%%%%%%%%%%%%%%%%%%%%%%%%%%%%%%%%%%

%%%%%%%%%%%%%%
%Combining in a compact notation, we can write $N_\alpha = \sum_aU_{\alpha a}N_a$ where $\alpha = e, \mu$, the flavor label  and $a$ denotes the mass eigenstate label $a=1,2$. 
%Once either $N_{e,\mu}$ is produced in $W_R$ decay, this will evolve according to the mass eigenstates, after a time $t$, the amplitudes for $ A_{OSee}, A_{SSee},
%A_{OSe\mu}, A_{SSe\mu}, A_{OS\mu e}, A_{SS\mu e}, A_{OS\mu\mu}, A_{SS\mu\mu}$ can be written from which one can get $ N_{OSee},  N_{SSee},
%N_{OSe\mu}, N_{SSe\mu}, N_{OS\mu e}, N_{SS\mu e}, N_{OS\mu\mu}, N_{SS\mu\mu}$.
%Using the compact notation above, the amplitudes for same (SS) and opposite sign (OS) dileptons can be written as:
%\begin{eqnarray}
%A_{SS\alpha\beta}~=~\sum_a U_{\alpha a}U^*_{\beta a} e^{-iE_a t-\frac{\Gamma_a t}{2}}\\ \nonumber
%A_{OS\alpha\beta}~=~\sum_a U_{\alpha a}U_{\beta a} e^{-iE_a t-\frac{\Gamma_a t}{2}}
%\end{eqnarray}

\begin{figure}[t]
\centering
\includegraphics[width = 0.6\textwidth]{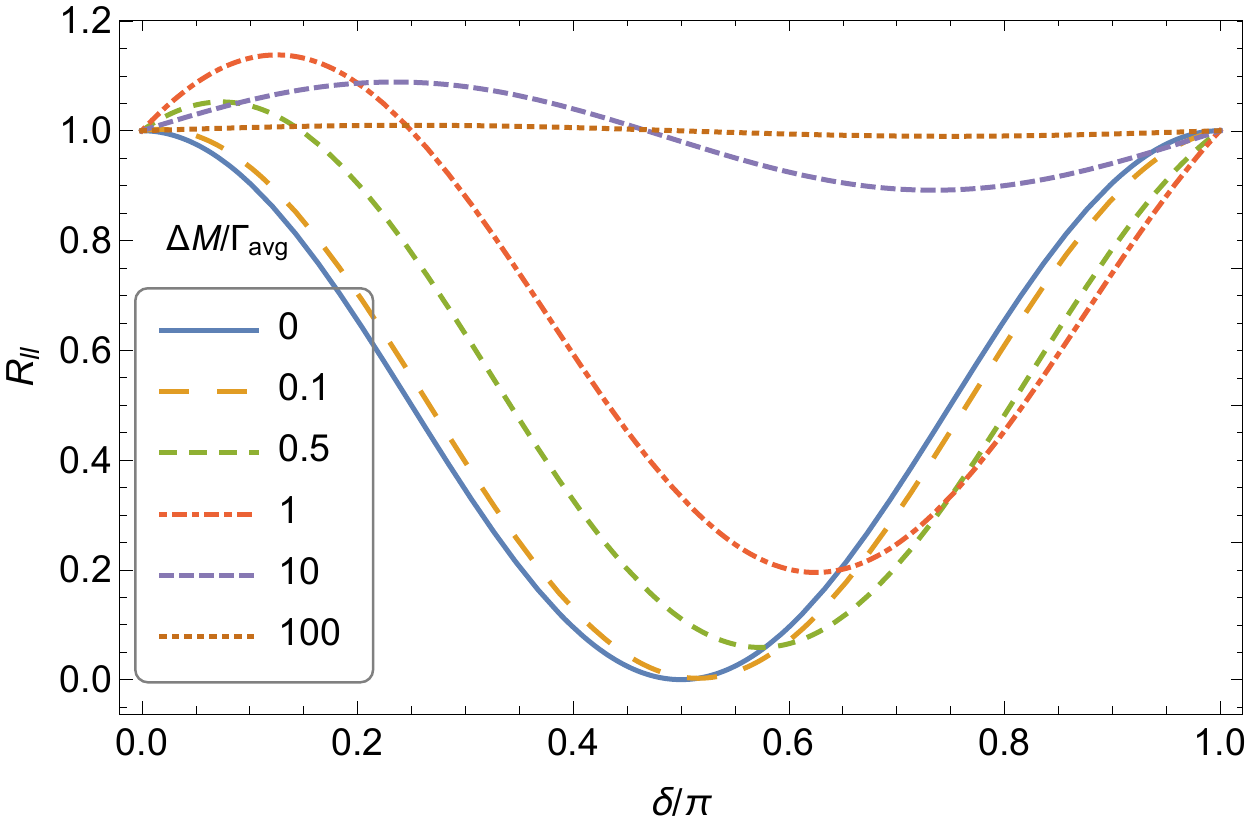}
\caption{The variation of $R_{\ell\ell}$ (with $\ell\ell=ee,\mu\mu$) as a function of the $CP$ phase $\delta$ for different values of $\Delta M/\Gamma_{\rm avg}$ in the TeV-scale left-right type-I seesaw model.} 
\label{fig:Rll-typeI}
\end{figure}

Expanding these equations out, we find that if there is no $CP$ phase i.e. $\delta=0$, we get
\begin{eqnarray}
 N_{{\rm OS},ee} \ = \ N_{{\rm SS},ee}\, ; \quad N_{{\rm OS},\mu\mu} \ = \ N_{{\rm SS},\mu\mu} \, ; \quad N_{{\rm OS}, e\mu} \ = \ N_{{\rm SS},e\mu} \, ,
 \end{eqnarray}
as expected for purely Majorana RHNs. However, in the presence of a non-zero $CP$ phase, we find  
 \begin{eqnarray}
 N_{{\rm OS},\ell\ell} \ \neq \ N_{{\rm SS},\ell\ell} \, , \quad {\rm or} \quad R_{\ell\ell}\neq 1 \, ,
  \end{eqnarray}
as illustrated in Figure~\ref{fig:Rll-typeI}. We emphasize again that these arguments are true only if the two RHN states satisfy the coherence conditions~\eqref{eq:coh1} and \eqref{eq:coh2}.
%, i.e. $\Delta M^2 \leq E\Gamma_{W_R}$\cite{Akhmedov:2007fk}. 

Let us apply our findings to the special case where the RHN mass matrix is of the form $M_N=M\tau_1$ where $\tau_1$ is the first Pauli matrix. In this case $\theta=\pi/4$ and $\delta=\pi/2$. Also in this case, $\Gamma_1=\Gamma_2$ and $\Delta M=0$. Substituting this in Eq.~\eqref{Rll}, we get  $N_{{\rm SS},ee}=N_{{\rm SS},\mu\mu}=0$ and only $N_{{\rm SS},e\mu}\neq 0$ as we would expect from the structure of the RHN mass matrix.\footnote{However, in this special case, $R_{e\mu}$ is ill-defined, because $N_{{\rm OS},e\mu}=0$.}

In Figure~\ref{fig:Rll-typeI}, we show the variation of $R_{\ell\ell}$ (for $\ell\ell=ee,\mu\mu$) as a function of the $CP$ phase $\delta$ for different values of $\Delta M/\Gamma_{\rm avg}$.  As for the RHN decay widths, we have used the three-body decay widths of $N_\ell\to W_R^*\ell \to q\bar{q}'\ell$ (see Appendix~\ref{sec:app}). For numerical purposes, we have chosen a fixed value of $M_{W_R}=5$ TeV and $M_{N_1}=500$ GeV, but our main results are independent of the choice of the exact mass values, as long as $M_{N_{1,2}}< M_{W_R}$, which is anyway required from vacuum stability arguments~\cite{Mohapatra:1986pj, Maiezza:2016bzp}. 

We find that for $\delta=0$ and $\pi$ (i.e. no $CP$ violation), $R_{\ell\ell}=1$, as discussed above. But for $\delta=\pi/2$ and $\Delta M=0$, $R_{\ell\ell}=0$, i.e. there is a completely destructive interference between the two RHN mass eigenstates in the SS channel. The degree of interference decreases rapidly as we increase $\Delta M$ and as $\Delta M$ becomes much larger than $\Gamma_{\rm avg}$, there is virtually no interference, leading to the limit $R_{\ell\ell}\to 1$, as expected for purely Majorana RHNs. In the intermediate range of $\Delta M/\Gamma_{\rm avg}$, we have $R_{\ell\ell}>1$, i.e. enhanced SS signal even compared to the purely Majorana case, for certain choices of the $CP$ phase $\delta$, when the constructive interference is maximum in Eqs.~\eqref{eq:SSee} and \eqref{eq:SSmumu}. Note here that both $ee$ and $\mu\mu$ channels lead to almost identical predictions for the ratio $R_{\ell\ell}$ because for $\Delta M\ll M$ in Eq.~\eqref{eq:theta}, $\theta\approx \pi/4$, so $c_\theta\simeq s_\theta\simeq 1/\sqrt{2}$.

 \section{General Inverse Seesaw Case} \label{sec:inv}

We start this section by briefly reviewing the inverse seesaw extension of the left-right symmetric model. The model is based on the gauge group $SU(3)_c\times SU(2)_L\times SU(2)_R \times U(1)_{B-L}$ gauge group~\cite{LR1, LR2, LR3} with scalar sector consisting of two $SU(2)$ doublets $\chi_R^{}(1,1,2,+1)$, together with a bidoublet $\phi(2,2,0)$ and a $B-L=2$ triplet $\Delta_R(1,3,+2)$, while the fermion sector contains not only the usual $SU(2)$ doublets of the left-right model i.e. $Q_L^{}(3,2,1,+\frac{1}{3})$, $Q_R^{}(3,1,2,+\frac{1}{3})$, $L^{}(1,2,1,-1)$ and $R^{}(1,1,2,-1)$, but also additional $SU(2)$ singlets $S_a$ (with $a=1,2,3$). Note that we are working with a model where parity symmetry breaking scale $M_P$ and the $SU(2)_R$ symmetry breaking scale $v_R$  are different with $M_P\gg v_R$~\cite{Chang:1983fu}. %\footnote{The collider analysis of the inverse seesaw case in left right model has been studied in \cite{Das:2016akd} from the trilepton final state.}
 
 To discuss the inverse seesaw in this model, we need the leptonic Yukawa couplings:
 \begin{eqnarray}
 {\cal L}_Y \ = \ h_l\bar{L}\phi R+h_\nu \bar{R}\chi_R S +f\bar{R}^C  \Delta_R R+\mu_s \bar{S}^CS+{\rm H.c.}
 \end{eqnarray}
 After symmetry breaking by the vacumm expectation values of the Higgs fields i.e. ${\rm Diag} \langle\phi \rangle=(\kappa, \kappa')$; $\langle \chi^0_R\rangle = \sigma_R$ and $\langle\Delta^0_R\rangle = v_R$, we get the neutral fermion mass matrix of the form:
%\newline 
 \begin{eqnarray}
{\cal M} \ &=& \ \left(\begin{array}{ccc} 0 & M_D & 0\\ M^T_D & \mu_R & M_N\\
0 & M^T_N & \mu_S \end{array}\right)
\label{eq:4.2}
\end{eqnarray}
where $M_D=h_l\sqrt{\kappa^2+\kappa'^2}$,  $\mu_R=fv_R$ and $M_N=h_\nu \sigma_R$.\footnote{Note that the left-right symmetry does not allow a (1,3) entry in Eq.~\eqref{eq:4.2}, which would otherwise lead to the linear seesaw~\cite{Barr:2003nn,Malinsky:2005bi}.} It leads to the formula for light neutrino mass matrix at tree-level:\footnote{The $\mu_R$ term leads to unavoidable one-loop corrections to the light neutrino mass matrix, but for a given $\mu_R$, we can carefully choose $\mu_S$ so that the light neutrino oscillation data is always satisfied~\cite{Dev:2012sg}.}
\begin{eqnarray}
M_\nu \ = \ (M_DM_N^{-1}) \mu_S(M_DM_N^{-1})^T \, .
\label{eq:4.3}
\end{eqnarray}

This is the inverse seesaw mechanism at work for the most general case where each entry in Eq.~\eqref{eq:4.3} is a $3\times 3$ matrix corresponding to three flavors. For simplicity below we consider a single family version of this matrix to illustrate our discussion of SS and OS dilepton plus two jets in $pp$ collision. We note that this analysis can be applicable to realistic situation with flavor in the following way:  Consider the case when $M_D, M_N, \mu_R$ are all diagonal $3\times 3$ matrices and let all neutrino flavor mixings reside in the $\mu_S$ matrix. This is the so-called flavor-diagonal scenario. Proper choice of the $\mu_S$ matrix can explain the observed neutrino oscillation results but since each element of the $\mu_S$ matrix is very small compared to other matrices in the problem i.e. $\mu_R, M_N$, they will not affect our conclusions about the ratio $R_{\ell\ell}$ for each flavor. Of course one could also consider flavor structures in $\mu_R$ and/or $M_N$. The analysis is then more complicated and we do not consider it here.\footnote{For the case of non-diagonal $M_D$, one should also make sure to satisfy the experimental constraints from lepton flavor violating processes such as $\mu \to e\gamma$.} %The FD scenario is relatively easy to test at the colliders. The detailed collider analyses is beyond the scope of the paper and will be considered in a follow up work.}} 

\section{$R_{\ell\ell}$ in the Inverse Seesaw Case}\label{sec:inv2}
%Inverse seesaw matrix for the most general case in the $(\nu, N, S)$ basis is given by:
%\begin{eqnarray}
%M_\nu~=~\left(\begin{array}{ccc} 0 & m_D & 0\\m^T_D & \mu_R & M_N\\
%0 & M^T_N &\mu_S\end{array}\right)
%\end{eqnarray}
%For simplicity, we consider the case of one fermion generation within the context of an left-right model.
%We assume a 5 TeV mass for $W_R$ and $M_N\sim$ TeV or less. 
%We assume that the $W_R$ is produced in the pp-collision at LHC or higher energy/or luminosity pp collider. The $W_R$ then decays to $N$ which is part of the right handed $SU(2)_R$ doublet as $W_R\to \ell +N$. 
In order to study the final states with SS or OS dileptons, we consider a simplified yet realistic case where $\mu_R$ and $M_N$ are $2\times 2$ diagonal matrices so that all neutrino mixings arise from the matrix $\mu_S$, which does not have any effect on $R_{\ell\ell}$. We consider the eigenstates of the mass matrix~\eqref{eq:4.2}. We do this in stages and for the parameter domain where $M_N\gg \mu_R \gg M_D \gg \mu_S$, we can first diagonalize the lower $2\times 2$ matrix and get the following eigenstates with real eigenvalues
\begin{eqnarray}
{\cal N}_{1} \ &=& \ c_\alpha N+s_\alpha S \, ; \\
{\cal N}_{2} \ &=& \ i(-s_\alpha N+c_\alpha S) \, .
\end{eqnarray}
Using these we can rewrite the $W_R$-induced charged-current interactions as 
\begin{eqnarray}
{\cal L}_I \ = \ \frac{g}{\sqrt{2}}\bar{\ell}_R\gamma_\mu (c_\alpha {\cal N}_1+is_\alpha {\cal N}_2)W^{-,\mu}_R+\frac{g}{\sqrt{2}} (c_\alpha {\cal N}_1-is_\alpha {\cal N}_2)^TC\gamma_\mu\ell_RW^{+,\mu}_R
\label{eq:laginv}
\end{eqnarray}
To calculate the OS and SS event numbers, we need to discuss whether there is coherence between the decays of $\mathcal{N}_{1,2}$ when produced in $W_R$ decay as well as the requirement for maintaining coherence over the decay length of the RHNs. 

For the inverse seesaw case,  the discussions in Ref.~\cite{Akhmedov:2007fk} lead us to the same coherence condition in the emission as stated in Sec.~\ref{sec:coh} except in this case, $\Delta M$ denotes the mass difference between the $N$ and $S$ fermions.  The condition on parameters from coherence length considerations are different here since the Dirac Yukawa couplings which dominate the decay of $N,S$ states are expected to be much larger for inverse seesaw than the type I case. The decay length in this case is therefore much shorter than the type I case i.e. $L\sim \frac{12\pi}{h^2 M_N}$. Choosing $h\sim 0.1$ and $M_N\sim $ TeV, we get for $\Delta M\sim 100$ GeV, as compared to 1 GeV or so in the type I case.
%We take as the conditions for coherence the requirement that the uncertainty in the mass square of the two heavy neutrino eigenstates $\sigma_{m^2}$ to satisfy the condition $\sigma_{m^2} \geq \Delta M^2\equiv |M^2_1-M^2_2|$. The $\sigma_{m^2}$ in this case is estimated to be $2\sqrt{2} E\Gamma_{W_R}$ where $E$ is the energy of the heavy neutrino eigenstates and $\Gamma_{W_R}$ is the width of the $W_R$ which causes the uncertainly in the  energy of the produced heavy state.
%In this case we find that coherence occurs when $\Delta M^2\leq E\Gamma_{W_R}$, which is satisfied for a TeV-scale $N$ when the mass difference between the states is less than 100 GeV. We have already estimated that $\Gamma_{W_R}\sim (g^2/12\pi)M_{W_R}$}. 
%When there is coherence, the  contributions from the two states interfere and their amplitudes must be added. When there is no coherence, the OS and SS terms are to considered separately and that means we will always get the ratio of SS to OS events $R_{\ell\ell}$ to be equal to one.

When the coherence condition is satisfied, recalling that the first term in the Lagrangian \eqref{eq:laginv} is responsible for the production of OS events and second one for SS events, we can write the amplitudes for OS and SS events as follows:
\begin{eqnarray}
A_{\rm OS}(t) \ &=& \ c^2_\alpha e^{-iE_1t-\frac{\Gamma_1}{2}t}+s^2_\alpha e^{-iE_2t-\frac{\Gamma_2}{2}t} \, , \nonumber \\
A_{\rm SS}(t) \ &=& \ c^2_\alpha e^{-iE_1t-\frac{\Gamma_1}{2}t}-s^2_\alpha e^{-iE_2t-\frac{\Gamma_2}{2}t} \, .
\end{eqnarray}
We approximate $E_{1,2}\simeq M_{1,2}\pm \Delta M/2$ as before and use the expression in Eq.~\eqref{Rll} to obtain for the OS and SS events respectively 
%taking the number of for OS and SS to be (choosing $T\sim \frac{1}{\Gamma_{1,2}}$),
%\begin{eqnarray}
% N_{OS}=\frac{1}{T}\int^\infty_0 dt  |A_{OS}(t)|^2;\\ \nonumber
% N_{SS}=\frac{1}{T}\int^\infty_0 dt  |A_{SS}(t)|^2
 %\end{eqnarray}
% we get for the ratio of SS to OS leptons as
 \begin{align}
N_{\rm OS} \ & = \  \Gamma_{\rm avg}    \left[      \frac{c_\alpha^4}{\Gamma_1}+\frac{s_\alpha^4}{\Gamma_2}-\frac{c_\alpha^2 s_\alpha^2(\Gamma_1+\Gamma_2)}{\frac{(\Gamma_1+\Gamma_2)^2}{4}+(\Delta M)^2}\right], \label{eq:OS} \\
N_{\rm SS} \ & =  \ \Gamma_{\rm avg}    \left[   \frac{c_\alpha^4}{\Gamma_1}+\frac{s_\alpha^4}{\Gamma_2}+\frac{c_\alpha^2s_\alpha^2(\Gamma_1+\Gamma_2)}{\frac{(\Gamma_1+\Gamma_2)^2}{4}+(\Delta M)^2} \right] \, . \label{eq:SS}
\end{align}
% R_{\ell\ell} \ = \ \frac{\frac{c^4}{\Gamma_1}+\frac{s^4}{\Gamma_2}-\frac{c^2s^2(\Gamma_1+\Gamma_2)}{\frac{(\Gamma_1+\Gamma_2)^2}{4}+\Delta M^2}}{\frac{c^4}{\Gamma_1}+\frac{s^4}{\Gamma_2}+\frac{c^2s^2(\Gamma_1+\Gamma_2)}{\frac{(\Gamma_1+\Gamma_2)^2}{4}+\Delta M^2}}
 %\label{Rll2}
 %\end{eqnarray}
Using the RHN decay widths  given in Appendix~\ref{sec:app}, we have plotted in Fig.~\ref{fig:Rll-ISS} the ratio $R_{\ell\ell}=N_{\rm SS}/N_{\rm OS}$ as a function of $\mu_R$ for different RHN masses. Here we have chosen a fixed value of  $M_{W_R}=5$ TeV for illustration. We find that smaller values of $\mu_R$ favors the OS signal whereas higher values of $\mu_R$ favor the SS signal. For lower values of $M_N$, the range of $\mu_R$ increases where $R_{\ell\ell}\to 1$.

%%%%%%%%%%%%%%%%%%%%%%%%%%%%%%%%%
\begin{figure}[t]
\centering
\includegraphics[width = 0.6\textwidth]{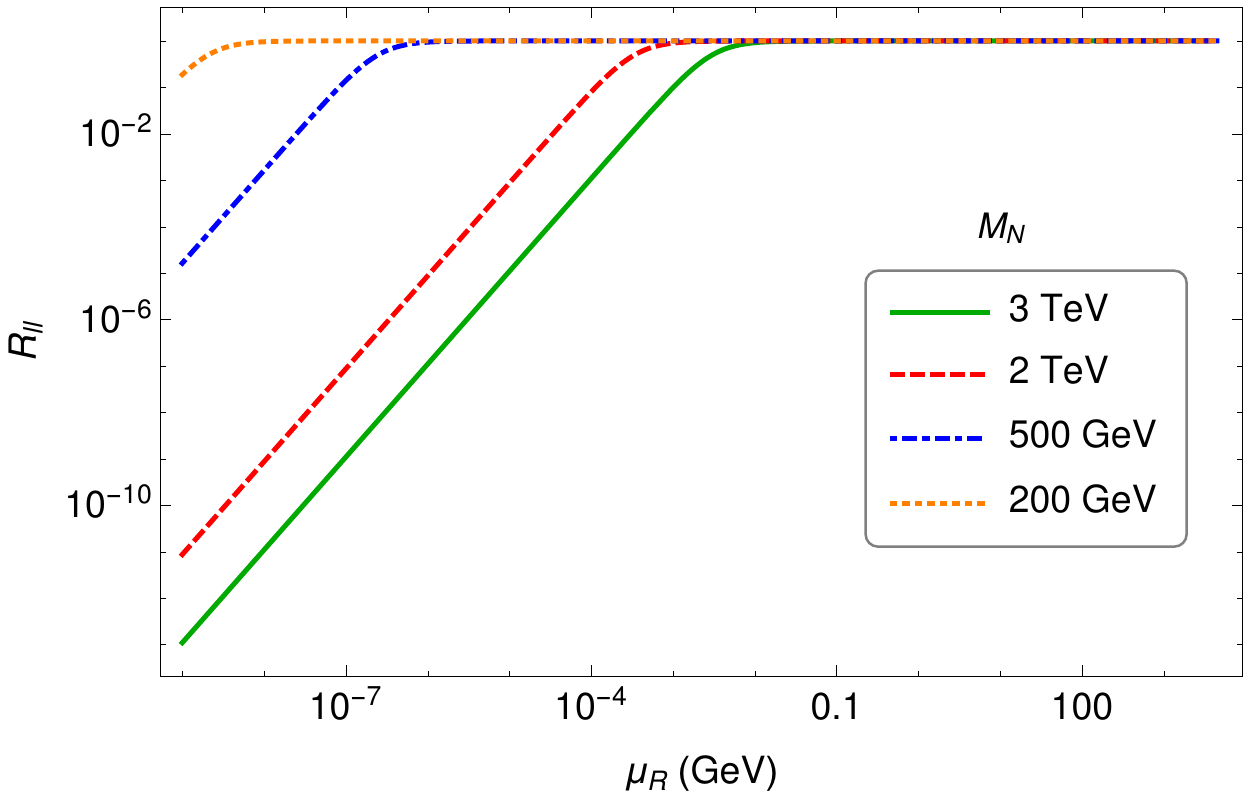}
\caption{The variation of $R_{\ell\ell}$ as a function of $\mu_R$ for different values of $M_N$ in the left-right model for the inverse seesaw case.} 
\label{fig:Rll-ISS}
\end{figure}
 
Now we can look at three special cases:
 
\noindent {\bf Case (i): $\mu_R=0$}: This is the case which has been considered in Refs.~\cite{Anamiati:2016uxp, Antusch:2017ebe}. In this case for a TeV $M_N$, fitting neutrino mass scale requires that $\mu_S\leq $ keV. This means that $\Delta M \sim \mu_S\sim $ keV and the coherence condition is very well satisfied. Furthermore, in this case $c_\alpha=s_\alpha = \frac{1}{\sqrt{2}}$. Using the fact that we have also $\Gamma_1=\Gamma_2$, we get from Eqs.~\eqref{eq:OS} and \eqref{eq:SS}, that
 \begin{eqnarray}
 R_{\ell\ell} \ = \ \frac{(\Delta M)^2}{2\Gamma^2+(\Delta M)^2}
 \end{eqnarray}
 in agreement with the result in Ref.~\cite{Anamiati:2016uxp}. Note that for TeV-scale $M_{N}$, typically $\Gamma\sim 10-100$ keV and $\Delta M\sim 1$ keV, leading to $R_{\ell\ell}\lesssim 1\%$. Thus to get large $R_{\ell\ell}$ in inverse seesaw models, one must include the effect of $\mu_R$. 
 
% {\bf \Large Please check all these numbers and formulae}
 
\noindent {\bf Case (ii) $\mu_R\ll M_N$}: In this case, in general $\alpha$ is different from $\pi/4$ and we do not expect $\Gamma_1$ and $\Gamma_2$ to be equal. If we assume that $\Gamma_1\sim \Gamma_0 c^2_\alpha$ and $\Gamma_2\sim \Gamma_0 s^2_\alpha$, we get 
 \begin{eqnarray}
 R_{\ell\ell} \ = \ \frac{\cos^2 2\alpha+\frac{4(\Delta M)^2}{\Gamma^2_0}}{1+\sin^22\alpha+\frac{4(\Delta M)^2}{\Gamma^2_0}} \, .
 \end{eqnarray}
 For the case when $\frac{4(\Delta M)^2}{\Gamma^2_0}\ll 1$, it reduces to the formula in Ref.~\cite{Dev:2015kca}. In this case, $R_{\ell\ell}$ can be significant; see Figure~\ref{fig:Rll-ISS}. 

\noindent{\bf Case (iii): Hierarchical masses i.e. $\mu_R \gg M_N$:} In this case, the two eigenstates ${\cal N}_{1,2}$ have a large mass difference i.e. $(\Delta M)^2\gg \Gamma^2_{1,2}$. In this case, there is no coherence and we have therefore $R_{\ell\ell}=1$ as in the type-I seesaw case since the two Majorana eigenstates both lead to equal number of SS and OS dilepton states.

%%%%%%%%%%%%%%%%%%%%% FIGURE 
%%%%%%%%%%%%%%%%%%%%%%%%%%%%%%%%%%%%%%%%%%%%%%%%%%%%%%%%%%%%%%

\section{Conclusion} \label{sec:con}

We show that in generic TeV scale $W_R$ models for type I and  general inverse seesaw models, the ratio  $R_{\ell\ell}$, of the number of same sign ($N_{\rm SS}$) and opposite sign ($N_{\rm OS}$)  dilepton states need not be the same when summed over different flavors, contrary to general expectations. This can happen when there is a high degree of degeneracy between the RHNs produced in $W_R$ decay. The degree of degeneracy depends on whether it is type I or inverse seesaw case, and is determined by the coherence condition which in turn depends on the magnitude of the Dirac Yukawa couplings in the theory. For generic choice of parameters, in the first case, the degeneracy has to be at the level of one part in a thousand for TeV scale RHNs whereas in the case of inverse seesaw, it can be a factor of ten or less. Thus observation of the ratio $R_{\ell\ell}$ can in principle, allow us to probe  the deeper  structure of the RHN mass matrix in the type I seesaw case and the $(N,S)$ sector mass matrix in the inverse seesaw case. We find that in the case of type I seesaw, one needs $CP$ violation to get $R_{\ell\ell}$ different from one, whereas for the inverse seesaw, it is the parameter $\mu_R$ which governs $R_{\ell\ell}$. We believe that the connection between $R_{\ell\ell}\neq 1$ and near degeneracy of RHN states is already an important conclusion, since it is known that  low scale leptogenesis in TeV scale seesaw models already requires near degeneracy of RHN states.

Our main goal in this work was to derive the analytic results for $R_{\ell\ell}$ in the singlet seesaw scenario, and to show as a proof of principle that it can be different from 1 in the parameter space relevant for the LHC. This result is valid irrespective of the details of the collider simulation of the OS and SS events, with their respective signal and background efficiencies, which can be done in a straightforward manner for any given benchmark point following the existing experimental analyses; see e.g. Ref.~\cite{Khachatryan:2014dka}. 
%we believe is a bit premature at this stage. 
Also in the case of inverse seesaw, we have ignored detailed flavor effects, since our goal has been merely to illustrate an interesting phenomenon involving lepton {\it number} violation. A detailed collider analysis (including flavor effects) is a bit premature at this stage and might be more appropriate in scrutinizing the different seesaw models, only if there is a statistically significant observation of dilepton plus two jet signal (either SS or OS) in the future.\footnote{It might be noted here that CMS had reported a local $2.8\: \sigma$ $eejj$ excess, mostly in the OS dilepton events, in the  $\sqrt s=8$ TeV LHC data~\cite{Khachatryan:2014dka},  which led to a flurry of theoretical interpretations, but this was not confirmed in the $\sqrt s=13$ TeV data~\cite{CMS:2017uoz, CMS:2017ilm}.}

\section*{Acknowledgement} We gratefully acknowledge the local hospitality provided at the ACFI workshop on `Neutrinos at the High-Energy Frontier' at UMass, Amherst, where part of this work was done. R.N.M. was supported by the US National Science Foundation under Grant No. PHY1620074. 

\appendix
\section{Partial Decay Widths of $N$} \label{sec:app}
In the left-right model, the RHN has three-body decays through an off-shell $W_R$ (for $M_N<M_{W_R}$): $N_\ell\to W_R^*\ell \to \ell q\bar{q}^{\prime}$.  This is in addition to the usual two-body decay modes of the RHN: $N\to W\ell, \, Z\nu, \, h\nu$, induced by its mixing with the light neutrinos. In this analysis, we choose the region of parameter space where the light-heavy neutrino mixing is small enough to ensure that the three-body decay is dominant over the two-body one~\cite{Chen:2013fna}.  
%However, in the Such partial decay modes are proportional to the modulus square of the the light-heavy mixing parameter $|V_{\ell N}|^2$. In this analysis
%we chose such a region where $|V_{\ell N}|^2$ is too low such that the decay of $N$ will be dominant through $W_R$. As $M_{W_R} \gg M_N$ therefore the decay modes of $N$ through the $W_R$ will be three body. The corresponding decay widths can be written as

For light-quark final states, the corresponding three-body decay width is given by~\cite{Gluza:2015goa, Das:2016akd}
\begin{equation}
 \Gamma(N \to q\bar{q}^\prime \ell ) \ = \  
\frac{g_R^4}{2048 \pi^3} M_{N}  \frac{12}{x}\left[1-\frac{x}{2}-\frac{x^2}{6}+\frac{1-x}{x}\ln (1-x)\right] ~,
\end{equation}
with $x = M_{N}^2/M_{W_R}^2$. Here we neglect the SM quark and lepton masses. For the $N \to
t\bar{b}  \ell$ decay channel, we have~\cite{Dobrescu:2015jvn, Das:2016akd} 
\begin{align}
 \Gamma (N \to \overline{b} t \ell) \ = \  
\frac{g_R^4}{2048 \pi^3} M_{N} F_t(x, y) ~,
\end{align}
where
\begin{align}
 F_t(x, y) &\ = \ \frac{12}{x} \biggl[
(1-y) -\frac{x}{2}(1-y^2) -\frac{x^2}{6}\left(
1-\frac{3}{2}y + \frac{3}{2}y^2 -y^3 
\right) \nonumber \\[3pt]
&-\frac{5x^3y}{8}(1-y^2) +\frac{x^4y^2(1-y)}{4}
-\frac{x^3y^2}{4}(4+x^2y)\ln y \nonumber \\
&+ \frac{1-x}{x} \ln \left(\frac{1-x}{1-xy}\right)
\left\{1-\frac{xy}{4}\left[4+x+x^2-x^3 y^2 (1+x)\right]\right\}
\biggr] ~,
\end{align}
with $y=m_t^2/M_N^2$ and $m_t$ is the top mass. These decay widths have been used in our numerical analysis for $R_{\ell\ell}$ (see Figures~\ref{fig:Rll-typeI} and \ref{fig:Rll-ISS}) with a benchmark value of $M_{W_R}=5$ TeV and $M_N=500$ GeV. As long as $M_N\ll M_{W_R}$, the actual values of these masses do not affect our final results.

\end{document}